\documentstyle[12pt]{article}

\begin{document}
\sloppy
\begin{flushright}{TIT/HEP-394\\ May '98}\end{flushright}
\vskip 1.5 truecm
\centerline{\large{\bf On Superpotential and BPS domain wall }}
\centerline{\large{\bf in SQCD and MQCD}}
\vskip .75 truecm
\centerline{\bf Tomohiro Matsuda
\footnote{matsuda@th.phys.titech.ac.jp}}
\vskip .4 truecm
\centerline {\it Department of physics, Tokyo institute of technology}
\centerline {\it Oh-okayama, Meguro,Tokyo 152-8551, Japan}
\vskip 1. truecm
\makeatletter
\@addtoreset{equation}{section}
\def\theequation{\thesection.\arabic{equation}}
\makeatother
\vskip 1. truecm
\begin{abstract}
\hspace*{\parindent}
We examine the decoupling properties of the N=2 SQCD vacua when the
adjoint mass term is turned on and then the N=1 limit is taken.
The BPS domain wall tension in N=1 MQCD and SQCD is also examined.
The correspondence of the MQCD integrals with the superpotential
and the gaugino condensate is shown.
\end{abstract}
\newpage
\section{Introduction}
\hspace*{\parindent}
Supersymmetric gauge theories have the special property that certain
quantities are holomorphic and can often be obtained exactly.
The exact results so obtained provide us an insight into the dynamics of
strongly coupled gauge theories, revealing a variety of interesting
phenomena.

In the case of $SU(N_{c})$ N=1 supersymmetric QCD with
$N_{f}$ flavors(SQCD), the exact superpotential is calculated
for $N_{f}<N_{c}$ and the holomorphic constraint is obtained
for $N_{f}=N_{c}$.
These theories are sometimes related to the N=2 derived N=1 SQCD
that is constructed by adding  the explicit mass term $\mu\Phi^{2}$
for the adjoint matter field $\Phi$.
Naively, N=2 derived SQCD seems to be connected smoothly to
ordinary N=1 SQCD when the decoupling limit $\mu\rightarrow\infty$
is taken.
In section 2, however, we show that this naive picture does not always
give a correct answer for the massless limit of N=2 and N=1 SQCD
because  $\mu\rightarrow\infty$ and $m_{f}\rightarrow 0$ do not
commute.
First we find a new (but naive) solution for the massive SQCD
that has the required properties of the N=1 limit
such as the $N_{c}$ degenerated vacua and the decoupling of
the extra $N_{c}-N_{f}$ vacua.
Introducing this solution that was missed in ref.\cite{hori}, we
find that there exist three
different types of  N=1 massless limits for N=2 massive SQCD,
corresponding to what limit we consider for  $\mu m_{f}$ and
$\mu m_{f}^{2}$.
These solutions are characterized by their discrete symmetries and
the decoupling properties.
Decoupling properties are discussed by means of the BPS bound for
the wall configuration.
Although we are not sure if the BPS domain wall can exist for these
configurations, the BPS bound is plausible as is discussed in
ref.\cite{shifman}

In section 3 we  examine  the MQCD calculation of the
BPS domain wall tension for N=1 massive theory
 and discuss its properties.
Our main concern is to examine the discrepancy between the MQCD brane
explanation and the SQCD domain wall derived from
the effective Lagrangian.
In this process, we find the integrals which correspond to  the
superpotential and the gaugino condensate.
These quantities are important when one considers the relation
between SQCD and MQCD.

\section{Field theory analysis}
\hspace*{\parindent}
In this section we consider a field theory analysis of
N=2 SQCD and examine  the vacuum structure
obtained by breaking the N=2 supersymmetry to N=1 by adding a
mass term for the adjoint
chiral multiplet.
We find a new (but naive)
solution  which has a smooth limit to  ordinary N=1 SQCD in the sense
that the discrete symmetry and the decoupling property are properly
satisfied.
Several types of the massless limits are also considered and characterized
by their discrete symmetries and the decoupling properties.

\underline{The discrete symmetry and the decoupling of the $N_{c}-N_{f}$
vacua}

First we consider the N=2 $SU(N_{c})$ gauge theory with
$N_{f}$ quark hypermultiplets in the fundamental representation.
In terms of N=1 superfields the vector multiplet consists of
a field strength chiral multiplet $W^{\alpha}$ and a scalar chiral
multiplet $\Phi$ both in the adjoint representation of the gauge group.
The N=2 superpotential takes the form
\begin{equation}
W=\sqrt{2}\tilde{Q}\Phi Q +\sqrt{2}m_{f}\tilde{Q}Q.
\end{equation}
To break the N=2 supersymmetry to N=1, a bare mass term is added
to the adjoint chiral multiplet $\Phi$
\begin{equation}
W=\sqrt{2}\tilde{Q}\Phi Q +\sqrt{2}m_{f}\tilde{Q}Q +\mu Tr(\Phi^{2}).
\end{equation}
When the mass $\mu$ for the adjoint chiral multiplet is increased
beyond $\Lambda_{N=2}$, the renormalization group flow below the
scale of  $\mu$ is the same as in N=1 SQCD with the dynamical
scale $\Lambda$ given by
\begin{equation}
\Lambda^{3N_{c}-N_{f}}=\mu^{N_{c}}\Lambda^{2N_{c}-N_{f}}_{N=2}.
\end{equation}
If the adjoint multiplet is much heavier than the dynamical scale,
we can first integrate out the heavy field $\Phi$ obtaining the
superpotential proportional to $\frac{1}{\mu}$
\begin{equation}
W_{\Delta}=\frac{1}{2\mu}\left[Tr(M^{2})-\frac{1}{N_{c}}(TrM)^{2}
\right].
\end{equation}
On the other hand, in N=1 SQCD with $N_{f}<N_{c}$, it is well known
that a superpotential is dynamically generated and takes the form
\begin{equation}
W_{ADS}=(N_{c}-N_{f})\left(\frac{\Lambda^{3N_{c}-N_{f}}}{detM}
\right)^{1/(N_{c}-N_{f})}.
\end{equation}
For large but finite $\mu$, we can expect that the effective
superpotential is just the sum of these terms,
\begin{eqnarray}
W_{eff}&=&W_{ADS}+W_{\Delta}\nonumber\\
&=&(N_{c}-N_{f})\left(\frac{\Lambda^{3N_{c}-N_{f}}}{detM}\right)^{1/(
N_{c}-N_{f})}+
\frac{1}{2\mu}\left[Tr(M^{2})-\frac{1}{N_{c}}(TrM)^{2}\right].
\end{eqnarray}
Extremizing the superpotential, one can find that the
vacuum structure of this superpotential is determined by
at most two different values of $m_{i}$ that appears in the diagonal part
of $M$.
The equation determining $m_{1}$ and $m_{2}$ is
\begin{eqnarray}
\label{moto_eq}
m_{i}^{2}-\frac{1}{N_{c}}\left[\sum^{N_{f}}_{l=1}m_{l}\right]m_{i}+\mu
m_{f}m_{i}=\mu\left(\frac{\Lambda^{3N_{c}-N_{f}}}{\Pi_{l}m_{l}}
\right)^{\frac{1}{N_{c}-N_{f}}}.
\end{eqnarray}
Subtracting the eq.(\ref{moto_eq}) for $i=1$ from the one for
 $i=2$, one obtains a equation
\begin{equation}
\label{sub}
(m_{1}-m_{2})\left[\left(1-\frac{r}{N_{c}}\right)m_{1}+
\left(1-\frac{N_{f}-r}{N_{c}}\right)m_{2}+\mu m_{f}\right]=0
\end{equation}
where $r$ denotes the number of $m_{1}$ that appears in the
diagonal part of $M$.
Although there are two types of solutions ($m_{1}\ne m_{2}$ or
$m_{1}=m_{2}$ corresponding to $r\ne 0$ or $r=0$)
 as seen from eq.(\ref{sub}), it is known\cite{hori} that
 $m_{1}\ne m_{2}$ and $r\ne 0$
is not suitable for the analysis of the N=1 limit.
As we are  concerned about the N=1 limit, we focus
our attention only on the solution
$m_{1}=m_{2}\equiv m$ and $r=0$.\footnote{Note that this naive
solution was misses in ref.\cite{hori}}
The equation for $m$ is now given by
\begin{equation}
\label{solution}
\frac{1}{\mu}
\left(1-\frac{N_{f}}{N_{c}}\right)m^{\frac{2N_{c}-N_{f}}{N_{c}-N_{f}}}
+m_{f}m^{\frac{N_{c}}{N_{c}-N_{f}}}-\Lambda^{\frac{3N_{c}-N_{f}}
{N_{c}-N_{f}}}=0.
\end{equation}
Taking the $\mu\rightarrow \infty$ limit, the contribution from
the first term becomes negligible
 and the $N_{c}$ solutions remain  finite.
The finite solutions are given by,
\begin{equation}
\label{sol_1}
m=\left(\frac{\Lambda^{3N_{c}-N_{f}}}{m_{f}^{N_{c}-N_{f}}}\right)
^{1/N_{c}}.
\end{equation}
Here the $N_{c}$ phase factors are implemented.
These solutions correspond to the well-known $N_{c}$ degenerated
vacua of N=1 SQCD.
In this paper,  we call these $N_{c}$ solutions by $m_{N=1}$ and
the remaining $N_{c}-N_{f}$ solutions by $m_{N=2}^{decouple}$.
From eq.(\ref{moto_eq}), it is easy to estimate the remaining
solutions $m_{N=2}^{decouple}$.
They are given by
\begin{eqnarray}
\label{sol_2}
m_{N=2}^{decouple}
&\sim& -(\mu m_{f})^{\frac{N_{c}-N_{f}}{N_{c}-N_{f}}}\nonumber\\
&\sim&-\mu m_{f}
\end{eqnarray}
where $N_{c}-N_{f}$ solutions are degenerated.
In the $\mu\rightarrow \infty$ limit, the latter solution
goes away to infinity and expected to be decoupled from $m_{N=1}$.
To ensure the decoupling of these vacua, it should be
important to examine the tension of the domain wall which interpolates
between $m_{N=1}$ and $m_{N=2}^{decouple}$.
The BPS bound for the tension of the domain wall that interpolates
between $m_{N=1}$ and $m_{N=2}^{decouple}$ is
\begin{eqnarray}
T_{D}&\geq& W_{eff}|_{m=m_{N=1}}-W_{eff}|_{m=m_{N=2}}\nonumber\\
&\sim&O(\mu)
\end{eqnarray}
which diverges in the N=1 ($\mu\rightarrow \infty$) limit.
The divergence of the domain wall tension indicates
 the decoupling of the $m_{N=2}$ solution.
This decoupling
property does not appear when one considers the solution given in
ref.\cite{hori}.
\footnote{If one solves the eq.(\ref{moto_eq}) for $m_{1}\ne m_{2}$
(see eq.(\ref{sub})) and assume that the obtained solution is also
applicable for $r=0$, one may find two solutions that looks very similar
to (\ref{sol_1}) and (\ref{sol_2}).
In this case, however, $m_{1}$ is not a physical solution
and does not appear in the Lagrangian even before one considers the
decoupling limit, since  $m_{1}$ is not a component of the meson
matrix $M$ when $r=0$.}

\underline{Massless limits}

Let us consider the massless limit of the solution and see how
the massless limit is realized in  N=1 and N=2 SQCD.
We are not interested in the massless theory itself but the symmetries
and the decoupling properties of the vacua that appears when one
takes the small mass limit.
We examine the N=1 ($\mu\rightarrow \infty, \Lambda=$fixed) and
massless ($m_{f}\rightarrow 0$) limit which have several branches
corresponding to what  value one chooses for
 $\mu m_{f}$ or  $\mu m_{f}^{2}$.

First let us consider the  $\mu m_{f}\rightarrow \infty$ limit.
In this limit the first term in eq.(\ref{solution}) is negligible,
 indicating the separation between two  solutions,
$m_{N=1}$ and $m_{N=2}^{decouple}$.
Although these two vacua are separated,
the decoupling is still not clear because
the tension of the domain wall which interpolates between
$m_{N=1}$ and $m_{N=2}$ is estimated as
\begin{eqnarray}
T_{D}&\geq& W_{eff}|_{m=m_{N=1}}-W_{eff}|_{m=m_{N=2}}\nonumber\\
&\sim& O(\mu m_{f}^{2})
\end{eqnarray}
and the value is not determined.
When $\mu m_{f}^{2}\rightarrow 0$, the bubbles of the
$m_{N=2}^{decouple}$ vacua can be
formulated in the $m_{N=1}$ vacua without costing any energy, thus
making the decoupling uncertain.
To ensure the decoupling of the extra vacua,
an additive requirement $\mu m_{f}^{2}\rightarrow \infty$ is needed.
When the $\mu m_{f}^{2}\rightarrow \infty$ limit is taken, $N_{c}$
vacua
are related by the discrete symmetry $Z_{N_{c}}$ while the other
$N_{c}-N_{f}$ vacua are decoupled.
The symmetry and the decoupling property show that
 this limit corresponds to the massless limit (to be more precise, what
we consider is the small mass limit) of the
ordinary N=1 SQCD.
Although both vacua run away as one takes the small mass limit,
one should not neglect these characteristic properties.

The next branch appears when one considers the limit
$\mu m_{f}\rightarrow 0$.
In this limit the second term in eq.(\ref{solution}) becomes
negligible and the $Z_{2N_{c}-N_{f}}$ degeneracy is
restored.
The resulting equation is precisely the one which appears
in the analysis of the $\mu\rightarrow\infty$
 limit of  N=2 {\bf massless} SQCD.
It is important  that the naive $\mu\rightarrow\infty$
 limit of N=2 massless SQCD
 lies at the opposite limit to the
massless limit of ordinary N=1 SQCD.

There is also a branch with $\mu m_{f}\rightarrow const$.
In this case, either $Z_{N_{c}}$ nor $Z_{N_{c}-N_{f}}$ symmetry is
restored and no decoupling is observed.
However, it is also easy to find that some other symmetry should be
restored.
Multiplying the eq.(\ref{solution}) by $\mu$ and
defining $x=m^{\frac{1}{N_{c}-N_{f}}}$, we can find
\begin{equation}
ax^{2N_{c}-N_{f}}+bx^{N_{c}}-\mu c=0
\end{equation}
where $a,b$ and $c$ are finite constants.
Since the last term depends on $\mu$, there should be
at least one vacuum that runs away.
On the other hand, to keep the coefficient of $x^{N_{c}}$
finite, some symmetry should be restored to ensure
the cancellation.
It is very easy to find the explicit example for $N_{c}=4, N_{f}=2$.

It should be useful to comment on  a remaining problem.
Because we do not know how to realize the new (and naive) $r=0$ solution
in MQCD, the relation between N=2 and N=1 MQCD is still unclear.
In this respect, the construction of the curve that realizes
the new $r=0$ solution seems an urgent task.

\section{SQCD and MQCD analysis of the BPS domain wall}
\hspace*{\parindent}
In this section we review and examine the calculation of BPS domain wall
tension in N=1 SQCD and MQCD to solve the discrepancy between these
two calculations.
We hope this provides us a key to understand more about the correspondence
between SQCD and  MQCD.
These analyses give us an important information on
the superpotential and the gaugino condensate.

Although the configuration of the five-brane for finite
 $\mu$ is already found in \cite{hori},
the equations followed by the curve does not contain
the $r=0$ solution of which we have discussed in the previous section.
Here we suspect the validity  of the  solution with
$m_{1}\ne m_{2}$ and $r=0$
(See the footnote in the previous section)
and consider only the N=1  curve that
can be obtained without referring to  the finite $\mu$ behavior.
According to  ref.\cite{veneziano}, the curve for N=1 SQCD is
\begin{eqnarray}
\label{curve1}
v&=&-\frac{\zeta}{\lambda}\nonumber\\
w&=&\lambda\nonumber\\
t&=&\lambda^{N_{c}-N_{f}}\left(\lambda-\frac{\zeta}{m_{f}}\right)^{N_{f}}.
\end{eqnarray}
The BPS domain wall tension  for this curve is already calculated
in ref.\cite{wall_SQCD}
 and given by:
\begin{equation}
\label{mqcd_wall}
T_{D}^{MQCD}=\zeta(2N_{c}-N_{f}).
\end{equation}
Here we follow the notations and the calculations
of ref.\cite{wall_SQCD}.
Although the  constant ambiguity is implemented,
$N_{c}$ and $N_{f}$ dependence  should be exact provided that the
physical observations  given by E.Witten\cite{Witten} are correct.

On the other hand, in the field theory analysis that was given in
ref.\cite{shifman}, the tension of the BPS domain wall is
calculated to be proportional
to the vacuum expectation value of the effective superpotential.
If this observation is correct, the tension should have only the
$N_{c}$ dependence and must not depend on $N_{f}$.

One way to solve this discrepancy between MQCD brane configuration and
the effective Lagrangian approach is to modify the original
integral for MQCD domain wall tension.
In ref.\cite{wall_SQCD}, a modification of the integral is proposed.
The main idea of ref.\cite{wall_SQCD} is to divide the integral
into two parts, namely the mesonic $w$-integral
\footnote{In the mesonic integral, $t$ is written by $w$ so that
 $W_{2}^{w}$ depends only on the mesonic variable $w$.
This is why we call $W_{2}^{w}$ mesonic integral.}
\begin{eqnarray}
\label{w}
W_{2}^{w}&=&\frac{1}{2\pi i}\zeta \int_{C_{w=\infty}}dt(w)/t\nonumber\\
&=&\zeta(\mbox{[No. of zeros] - [No. of poles of t(w)]})
\end{eqnarray}
and the $v$-integral
\begin{eqnarray}
\label{intv}
W_{2}^{v}&=&\frac{-1}{2\pi i}\zeta \int_{C_{v=\infty}}dt(v)/t\nonumber\\
&=&\zeta(\mbox{[No. of zeros] - [No. of poles of t(v)]})
\end{eqnarray}
and assume that the domain wall tension  $T_{D}$ should be written
as $T_{D}=W_{2}^{w}$.
\footnote{It should be noted that the definitions of the integral
(\ref{w}) and (\ref{intv}) are not new.
Integrating by parts and picking up surface terms, the integral
in ref.\cite{Witten} can  be divided into two surface terms
that correspond to (\ref{w}) and (\ref{intv}).}
Although this modification ($T_{D}=W_{2}^{w}+W_{2}^{v}\rightarrow
T_{D}=W_{2}^{w}$)
gives us the desired dependence on $N_{c}$,
it is clear that such a modification ruins the physical observations
 given by E.Witten in ref.\cite{Witten}.
The author of ref.\cite{wall_SQCD}
 claimed that such an ambiguity in defining the integral
may appear as the result of
the $\Sigma_{0}$ dependence of the Witten's definition and  proposed a
new integral that does not depend on $\Sigma_{0}$.
However, what he did to obtain the correct $N_{c}$
 dependence was to divide the {\bf new} integral
which he claimed to be $\Sigma_{0}$ {\bf independent}.
We should also remember that the domain wall can be constructed
as the brane configuration that interpolates $N_{c}$
degenerated vacua.
If such a configuration is considered, it does not depend on
the auxiliary $\Sigma_{0}$ curve.

The other way to understand this discrepancy between MQCD and
the effective Lagrangian approach is to examine
the field theory analysis.
To show that the MQCD  gives us the  correct answer,
first we try to explain why this discrepancy appears in these
two different approaches and then discuss more details on SQCD
calculation.

\underline{The basic idea revisited}

The key ingredient of the BPS domain wall
is the central extension of the N=1 superalgebra
\begin{eqnarray}
\label{algebra}
\{\overline{Q}_{\dot{\alpha}},\overline{Q}_{\dot{\beta}}\}&=&
        -2i(\sigma^{0})^{\gamma}_{\dot{\alpha}}\int d^{3}x
        \{\overline{D}_{\dot{\beta}}\overline{D}^{\dot{\delta}}
        J_{\gamma\dot{\delta}}\}_{\theta=0}.
\end{eqnarray}
The contribution to the central extension is due to the non-conservation
(classical or quantum anomaly) of the  current  multiplet $J$.
The lowest component of $J$ is the $U(1)_{R}$-current that is
broken classically (in Wess-Zumino model)
or by quantum anomaly ($U(1)_{R}$ anomaly which is induced
by gaugino).

The simplest example for the Kovner-Shifman domain wall is the
Wess-Zumino model with only a superfield $\Phi$ in which
 the presence of the central extension can be seen at the tree level.
In this case, the $U(1)_{R}$ symmetry is broken classically.
The Wess-Zumino Lagrangian in terms of a superfield $\Phi$
has the following form,
\begin{eqnarray}
L&=&\frac{1}{4}\int d^{4}\theta \Phi \overline{\Phi}+\left[
        \frac{1}{2}\int d^{2}\theta W(\Phi)+h.c.\right],\nonumber\\
W(\Phi)&=&\mu^{2}\Phi-\frac{\lambda}{3}\Phi^{3}.
\end{eqnarray}
When $\mu=0$, the Lagrangian is invariant under the $U(1)_{R}$ rotation
\begin{eqnarray}
  d\theta&\rightarrow& e^{-i\alpha}d\theta,\nonumber\\
  \Phi&\rightarrow&  e^{\frac{2i\alpha}{3}}\Phi.
\end{eqnarray}
However, if $\mu\ne0$, this $U(1)_{R}$ invariance is
broken and only the discrete part $Z_{2}^{R}$ persists,
\begin{eqnarray}
  d^{2}\theta&\rightarrow& -d^{2}\theta,\nonumber\\
  \Phi&\rightarrow&  -\Phi.
\end{eqnarray}
When $<\Phi>\ne 0$, the $Z_{2}^{R}$ symmetry is spontaneously broken
and the corresponding domain wall which interpolates between two
degenerate vacua
\begin{eqnarray}
\label{wz}
<\Phi>&=&\pm \mu/\sqrt{\lambda},\nonumber\\
<W>&=&\mp\frac{2}{3}\frac{\mu^{3}}{\sqrt{\lambda}}
\end{eqnarray}
appears.
From the definition of the $U(1)_{R}$ symmetry and the superpotential,
the superpotential must have the $U(1)_{R}$ charge that is broken if
the vacuum expectation value of $W$ becomes non-zero.
In this respect, the vacuum expectation value of the superpotential
parameterizes the classical (explicit) breaking of
the $U(1)_{R}$ current.
 The supersymmetric current multiplet $J$
has the anomaly of the following form,
\begin{eqnarray}
\overline{D}^{\dot{\alpha}}J_{\alpha\dot{\alpha}}&=&
\frac{1}{3}D_{\alpha}\left[3W-\Phi\frac{\partial W}{\partial \Phi}\right].
\end{eqnarray}
Substituting the anomaly equation into the central extension of the
superalgebra(\ref{algebra}), one obtains
\begin{eqnarray}
\{Q_{\alpha},Q_{\beta}\}&=&4(\vec{\sigma})_{\alpha\beta}
\int d^{3} x\vec{\nabla}\left[\overline{W}-\frac{1}{3}\overline{\Phi}
        \frac{\partial\overline{W}}{\partial\overline{\Phi}}\right]
        _{\overline{\theta}=0}
\end{eqnarray}
where the right hand side of the equation is related to the central
charge of the wall configuration therefore it represents  the surface
energy density of the wall.
It is apparent that the contribution is non-zero when the domain wall
interpolates the two different vacuum configurations of eq.(\ref{wz}).
The domain wall solution for such non-gauge theories
can easily be extended to more general
theories such as an effective theory of SQCD or some other complicated
theories.

Now let us discuss the SQCD with the superpotential.
As we have stated above, this theory has both the classical and
the quantum anomaly that contributes to the central extension.
The Lagrangian is
\begin{eqnarray}
L&=&\left[\frac{1}{4g^{2}}\int d^{2} \theta Tr{\cal W}^{2}+h.c.\right]
+\frac{1}{4}\int d^{2}\theta \left[\overline{Q}^{i}e^{V}Q^{i}+
\overline{\tilde{Q}}^{i}e^{V}\tilde{Q}^{i}\right].
\end{eqnarray}
In superfield notation the anomaly is
\begin{eqnarray}
\overline{D}^{\dot{\alpha}}J_{\alpha\dot{\alpha}}&=&
\frac{1}{3}D_{\alpha}\left\{\left[3W-\sum_{i}Q_{i}\frac{\partial W}
{\partial Q_{i}}\right]\right.\nonumber\\
&&\left.\left[\frac{3T(G)-\sum_{i}T(R_{i})}{16\pi^{2}}
Tr{\cal W}^{2}+\frac{1}{8}\sum_{i}\gamma_{i}Z_{i}\overline(D)^{2}
(\overline{Q}_{i}e^{V}Q_{i})\right]\right\}
\end{eqnarray}
where in SQCD,  $T(G)=N_{c}$ and $T(R_{i})=1$ for each flavor.
Substituting the anomaly equation into the superalgebra(\ref{algebra})
and using the Konishi anomaly equation, one gets
\begin{eqnarray}
\label{sqcd_cent}
\{\overline{Q}_{\alpha},\overline{Q}_{\beta}\}
&=&4(\vec{\sigma})_{\dot{\alpha}\dot{\beta}}
\int d^{3} x\vec{\nabla}\left[W-\frac{T(G)-\sum_{i}T(R_{i})}{16\pi^{2}}
Tr{\cal W}^{2}\right]_{\theta=0}.
\end{eqnarray}

For $N_{f}<N_{c}$ SQCD the domain wall solution is usually formulated for
the effective superpotential
\begin{eqnarray}
\label{eff}
W_{eff}&=&tr(m_{f}M)+c_{N_{c},N_{f}}\left(\frac{\Lambda^{3N_{c}-N_{f}}}
{detM}\right)^{N_{c}-N_{f}}
\end{eqnarray}
which looks like an extended Wess-Zumino model.
From eq.(\ref{sqcd_cent}), it is easy to understand
that there is always a contribution to the central charge
from the quantum $U(1)_{R}$ anomaly as far as
 gaugino condensation  takes place.
(Note that  the coefficient in eq.(\ref{sqcd_cent}) in front of
 gaugino condensation $Tr{\cal W}^{2}$
does not vanish for $N_{f}\ne N_{c}$.
Although the contribution from anomaly vanishes for $N_{f}=N_{c}$
massless SQCD, the
central charge does not vanish when the mass term is switched on
since  a contribution from  the superpotential
appears\cite{matsuda}.)
On the other hand, from eq.(\ref{eff}), one can see that  gaugino
 is already integrated out and thus
the quantum anomaly that is induced by gaugino is  decoupled
from the effective {\bf mesonic} Lagrangian(\ref{eff}).
In other words, the contribution from the quantum anomaly that appears
in the original Lagrangian is not included in the central charge
when it is derived from the {\bf mesonic} Lagrangian.
As we have mentioned in the last paragraph, $<W>$ parameterizes the
classical anomaly that  comes from the explicit
breaking of the $U(1)_{R}$ symmetry.
On the other hand, there is another contribution from the
vector superfield, which appears as a quantum anomaly of the
theory.
Adding these contributions, one can find
\begin{equation}
T_{D}^{SQCD}=|(2N_{c}-N_{f})\Lambda_{0}^{3}|
\end{equation}
where $\Lambda_{0}^{3}$ is the scale for the gaugino condensate.

In general, terms like $SlogS$ are required to include
the anomaly in the effective Lagrangian since the imaginary
part of the highest component of $S$ contains $F\tilde{F}$.
Although the glueball superfield $S$ is already integrated out,
 one may suspect that the shift of the
classical anomaly that is induced by the dynamical superpotential could be
identified with the shift by the quantum anomaly in the
original theory thus making the total central charge
$c\sim N_{c}\Lambda^{3}$.
To answer this question, it will be useful to think about the
theory where the dynamical superpotential is calculated without
using the effective Lagrangian.
In this respect, it shuld be useful to consider
SQCD with $N_{f}=N_{c}-1$.
In this case, one may calculate the dynamical superpotential
by using the  perturbative instanton calculation.
The dynamical superpotential appears as one instanton contribution
and is  calculated explicitly within the original (constituent) Lagrangian.
It shifts the vacuum expectation value of the superpotential
 and contributes to the central charge.
At the same time, quantum anomaly is calculated from the triangle
diagram or by using the well-known Fujikawa method.
It is apparent that these two (the shift from the dynamical superpotential
and the one from the quantum anomaly) appear as the independent
contributions
to the central charge.
Adding these contributions, one can find
\begin{equation}
T_{D}^{SQCD}=|(2N_{c}-N_{f})\Lambda_{0}^{3}|
\end{equation}
which agrees with the MQCD calculation when $\zeta$ in eq.(\ref{mqcd_wall})
is identified with $\Lambda^{3}_{0}$.
\footnote{
Some people may claim that there should be some unknown mechanism
that makes these two contributions identical, or claim that the
dynamical superpotential does not contribute to the central change
from unknown reason.
Here we respect MQCD calculation and do not consider
such arguments.}

\underline{MQCD correspondence}

As we have mentioned in the last paragraph, we may consider two independent
contributions (the classical and the quantum anomaly)
in the field theory analysis.
When one considers the MQCD integral for the domain wall tension,
mesonic $w$-integral (\ref{w}) is likely to
pick up the contribution from the mesonic field
as the mesonic Lagrangian picked up the classical anomaly.
In this sense, it is plausible that the contribution from
the mesonic Lagrangian eq.(\ref{eff}) appears as the $w$-integral
and is identified with the contribution from the classical anomaly.
On the other hand, the other contribution, namely the $v$-integral,
appears as the contribution from the quantum anomaly that is
decoupled in the mesonic picture.

We think it is very hard to prove this correspondence,
but it is very important to examine if these correspondence
 can survive in several cases.
Next we will examine them
by taking the massless and the decoupling  limits.

\underline{Massless limit}

In the  field theory language, the perturbative calculation at
the intermediate scale suggests
that the low energy effective Lagrangian for N=1 massless SQCD is
\cite{dine}
\begin{equation}
L_{eff}=L_{SU(N_{c}-N_{f})SYM}+L_{singlets}+O(1/M)
\end{equation}
where $O(1/M)$ denotes the higher order corrections that
interpolate between
the effective pure SYM sector ($L_{SU(N_{c}-N_{f})SYM}$)
and the singlet sector ($L_{singlets}$).
If one neglects the higher terms, the effective theory looks like
a pure SYM with the gauge group
$SU(N_{c}-N_{f})$  and likely to have the BPS domain wall whose
tension is $T_{D}=2(N_{c}-N_{f})\Lambda^{3}_{SU(N_{c}-N_{f})}$.
(Of course this result is not exact because the higher terms
are important when one discusses  the runaway.
We will find the gap appears also in MQCD.)

On the other hand, when massive SQCD is considered the wall tension
$T_{D}$ is always $T_{D}=(2N_{c}-N_{f})\Lambda^{3}_{0}$ for any value of
$m_{f}$.
This is because the expectation value of $W$ at its minimum does not
depend on the mass  $m_{f}$ as far as $m_{f}\ne 0$.

It should be important to examine these properties
 in the light of  MQCD and explain what happens in the massless limit.
As we have discussed in the last paragraph, we consider
the MQCD integral $W_{2}^{v}$ and $W_{2}^{w}$.
As is discussed above, $W_{2}^{w}$ corresponds
to  the mesonic superpotential and is always
$W^{w}_{2}=\zeta N$ as far as the variations of the parameters are smooth.
The curve (\ref{curve1}) in the $v$-picture is,
\begin{eqnarray}
\label{v_pic}
vw&=&-\zeta\nonumber\\
t&=&\zeta^{N_{c}}v^{-N_{c}}(m_{f}-v)^{N_{f}}.
\end{eqnarray}
When $m_{f}\rightarrow 0$ limit is taken, eq.(\ref{v_pic})
looks like a  SYM with the gauge group $SU(N_{c}-N_{f})$,
which coincides with the field theory analysis.
Although the domain wall tension $T_{D}=(2N_{c}-N_{f})\zeta$
remains the same
 as far as we keep $m_{f}\ne0$,
a discrepancy appears at  $m_{f}=0$.
When $m_{f}=0$ limit is taken, the $N_{f}$ zeros of $t(v)$ at
$v=m_{f}$ join the poles at $v=0$.
On the other hand, in the mesonic picture,  the $N_{f}$ zeros of $t(w)$
at $w=\zeta/m_{f}$ run away to infinity and finally
decouples from $W^{w}_{2}$.
As a result, the domain wall tension seems to have a gap at $m_{f}=0$,
which is the consequence of the runaway.
The gap found for the superpotential also indicates that
the theory is ill-defined in this limit.

It should be useful to examine further on $N_{c}=N_{f}$ in which
the vacua is well defined in the massless limit and
does not have a runaway behavior.
In this case, the curve (\ref{curve1}) becomes
\begin{eqnarray}
\label{curve3}
vw&=&-m_{f}\Lambda^{2}\nonumber\\
t&=&\left(w-\Lambda^{2}\right)^{N_{c}}
\end{eqnarray}
where we set $\Lambda^{2}=\frac{\zeta}{m_{f}}$.
It is easy to calculate the domain wall tension $T_{D}$ for
the curve (\ref{curve3}).
The result is,
\begin{equation}
T_{D}=N_{c}m_{f}\Lambda^{2}.
\end{equation}
In the field theory, the same result is already obtained
in ref.\cite{matsuda} by using the
holomorphic constraint.
As is discussed in the field theory in ref.\cite{matsuda},
the contribution comes solely from the mesonic integral.
In the field theory,  it is apparent that $T_{D}$ has a smooth massless
limit.
In MQCD, the curve (\ref{curve3}) in the massless limit becomes
\begin{eqnarray}
vw&=&0\nonumber\\
t&=&\left(\lambda-\Lambda^{2}\right)^{N_{c}}.
\end{eqnarray}
In this case we should think that the curve factorizes into
two different components,
\begin{eqnarray}
v&=&0\nonumber\\
t&=&(\lambda-\Lambda^{2})^{N_{c}}
\end{eqnarray}
and
\begin{eqnarray}
w&=&0\nonumber\\
t&=&\Lambda^{2N_{c}}.
\end{eqnarray}
Both give us the vanishing domain wall tension $T_{D}^{MQCD}$
,which is  consistent with the field theory analysis.
In this case, both $T_{D}$ and the superpotential
do not have a gap in this smooth massless limit.

\underline{Decoupling limit}

When the infinite mass limit is taken, a fundamental matter
superfield decouples.
If this decoupling is completed, the  effective
Lagrangian should be the N=1 SQCD with $N_{f}-1$ matter fields.
In this limit, the contribution from the quantum anomaly
$(N_{c}-N_{f})\Lambda^{3}_{0}$
should be replaced by $(N_{c}-N_{f}+1)\Lambda^{3}_{0}$.
Then the BPS domain wall tension is
changed from $T_{D}=(2N_{c}-N_{f})\Lambda^{3}_{0}$ to
$(2N_{c}-N_{f}+1)\Lambda^{3}_{0}$
because of  the gap in the quantum anomaly.
Is it possible to explain this  in the light of MQCD?

Here we consider the case one of the matter field($Q^{heavy}$)
acquires large mass($m^{heavy}$) and decouples.
In $w$-picture, when the decoupling limit is taken,
the zero at $\zeta/m_{f}^{heavy}$ join the
zeros at $w=0$ but the $w$-integral itself is not changed.
In $v$-picture,  the zero at $v=m_{f}^{heavy}$ decouples and then
the $v$-integral is changed.
As a result, $T_{D}$ becomes $T_{D}=(2N_{c}-N_{f}+1)\Lambda^{3}_{0}$.
This result agrees with the field theory analysis if $v$-integral
is identified with the contribution from the quantum anomaly.
The gap appears because the domain wall configuration cannot
survive the decoupling process.
Of course, the contribution from the superpotential ($W^{w}_{2}$) is
not changed.
This gives us an easy explanation why $T_{D}$ looks different in
the decoupling limit.

\underline{Summary}

In this section we analyzed the correspondence between the
SQCD $U(1)_{R}$ anomaly and the MQCD integrals.
From these analyses we can  confirm that $W^{w}_{2}$ and
$W^{v}_{2}$ should have  important physical meanings when they are
compared to the field theory.

\section{Conclusion and discussions}
\hspace*{\parindent}
In this paper we have found a new solution for $r=0$ and
discussed its properties.
First we examined the N=1 limit with massive quarks.
We have shown that the $Z_{N_{c}}$ degeneracy and the decoupling
of the extra $N_{c}-N_{f}$ vacua are realized in this limit.
Second we examined the N=1 and massless limit.
Keeping our attention on the symmetry and the decoupling
property of the vacua, we found three different limits
each of which are characterized by their discrete symmetries of the
degenerated
vacua and  decoupling properties.

We also examined the discrepancy between  MQCD and SQCD domain wall
calculation and found that the result given in ref.\cite{shifman}
should be modified to include the decoupled quantum anomaly.
In general, the integral for the MQCD domain wall surface energy
picks up two surface integrals which we call $w$ and $v$-integrals
in this paper.
Because the $w$-integral is given by the mesonic variable $w$,
it should be  natural to identify it with the mesonic superpotential.
If we may think that the $w$-integral corresponds to the superpotential,
the $v$-integral should correspond to the contribution
that comes from the quantum anomaly.
We examined this correspondence in several cases and found that
they are correctly realized.
These quantities must be important when one discusses the relation
between MQCD and SQCD.

\section{Acknowledgment}
It is a pleasure for me to express my gratitude to the people
in KEK and TIT for useful discussions on MQCD.

\end{document}